# Orthogonal polynomials derived from the tridiagonal representation approach

A. D. Alhaidari[†]

*Saudi Center for Theoretical Physics, P.O. Box 32741, Jeddah 21438, Saudi Arabia*

**Abstract**: The tridiagonal representation approach is an algebraic method for solving second order differential wave equations. Using this approach in the solution of quantum mechanical problems, we encounter two new classes of orthogonal polynomials whose properties give the structure and dynamics of the corresponding physical system. For a certain range of parameters, one of these polynomials has a mix of continuous and discrete spectra making it suitable for describing physical systems with both scattering and bound states. In this work, we define these polynomials by their recursion relations and highlight some of their properties using numerical means. Due to the prime significance of these polynomials in physics, we hope that our short expose will encourage experts in the field of orthogonal polynomials to study them and derive their properties (weight functions, generating functions, asymptotics, orthogonality relations, zeros, etc.) analytically.

## I. Introduction

The "*Tridiagonal Representation Approach* (TRA)" is an algebraic method for solving the quantum mechanical wave equation [1-5]

$$\left[-\frac{\hbar^2}{2M}\frac{d^2}{dx^2}+V(x)-E\right]\psi(E,x)=0, \tag{1}$$

for a class of potentials *V*(*x*) that is larger than the well-known conventional class (*M* is the effective mass of the model). In the TRA, the wavefunction $\psi(E,x)$ in configuration space for a given energy *E* is written as a bounded sum over a complete set of square integrable functions, $\{\phi_n(x)\}$. That is, we write $|\psi(E,x)\rangle=\sum_n f_n(E)|\phi_n(x)\rangle$, where $\{f_n(E)\}$ is an appropriate set of expansion coefficients. The basis functions are chosen such that the matrix representation of the wave operator is tridiagonal. This is possible only for a finite number of special potential functions. Nonetheless, the class of exactly solvable potentials in this approach is larger than the well-known conventional class. Recently, we made a short review of this approach and gave a partial list of these new solvable potentials (see, for example, Table 1 below) [6]. The tridiagonal requirement maps the wave equation (1) into a three-term recursion relation for the expansion coefficients. Subsequently, the recursion relation is solved exactly in terms of orthogonal polynomials $\{P_n(\varepsilon)\}$, where $\varepsilon$ is some proper function of the energy and $f_n(E)=$

---

[†] Present Address: 2403 Ahmed Bin Bilal Street, Al-Muhamadiyah 2, Jeddah 23624-6446, Saudi Arabia

$f_0(E)P_n(\varepsilon)$. Completeness of the basis and normalization of the wavefunction shows that $f_0^2(E)$ is the positive weight function associated with these energy polynomials [7-9]. The asymptotics ($n \to \infty$) of these polynomials all have the same general form

$$P_n(\varepsilon) \approx n^{-\tau} A(\varepsilon) \cos\left[ n^{\xi}\,\theta(\varepsilon) + \varphi(\varepsilon)\log n + \delta(\varepsilon) \right], \tag{2}$$

where $\tau$ and $\xi$ are real positive constants that depend on the particular polynomial. The studies in [7-12] show that $A(\varepsilon)$ is the scattering amplitude and $\delta(\varepsilon)$ is the phase shift. Both depend on the energy and physical parameters of the corresponding system. Bound states, if they exist, occur at energies $\{E(\varepsilon_m)\}$ that make the scattering amplitude vanish, $A(\varepsilon_m) = 0$. The size of this energy spectrum is either finite or infinite. Therefore, all physical information, structural and dynamical, about the system are given by the properties of these polynomials. Therefore, deriving these properties analytically is of prime importance. Such properties include the weight function, generating function, asymptotics, orthogonality relation, zeros, differential or difference relations, etc. Moreover, with each polynomial that has a continuous spectrum there is an associated discrete version that has an infinite or finite spectrum. These discrete polynomials constitute the expansion coefficients for the pure bound states.

In Ref. [6], we considered two classes of problems. One class is properly described in the "Laguerre basis" while the other is treated in the "Jacobi basis". In the Laguerre class, the basis functions are written as follows

$$\phi_n(x) = A_n y^{\alpha} e^{-\beta y} L_n^{\nu}(y), \tag{3}$$

where $y = y(x)$, $L_n^{\nu}(y)$ is the Laguerre polynomial of degree $n$ in $y$ and $A_n = \sqrt{\frac{\Gamma(n+1)}{\Gamma(n+\nu+1)}}$. The dimensionless real parameters $\{\alpha, \beta, \nu\}$ are obtained by the requirements of square integrability and tridiagonalization. The dimensionless variable $y$ is a function of the physical configuration space coordinate $x$ such that $y(x) \ge 0$, $\frac{dy}{dx} = \lambda y^a e^{by}$ and $2\beta = 1 + b$. The real dimensionless constants $a$ and $b$ are fixed for each choice of coordinate transformation associated with a given problem. Moreover, the length scale parameter $\lambda$ has an inverse length dimension (see Appendix A in Ref. [6] for details). The two remaining basis parameters $\alpha$ and $\nu$ are fixed by the physical parameters of the potential. The second order linear differential equation associated with all solvable problems in this class has the following form in the atomic units $\hbar = M = 1$

$$\left[ y\frac{d^2}{dy^2} + \left(a + by\right)\frac{d}{dy} + A_+ y + \frac{A_-}{y} \right]\psi(y) = \gamma\,\psi(y). \tag{4}$$

The corresponding potential functions in this class all have the following general structure [6]

$$V(x) = y^{2a} e^{2by}\left(V_0 + V_1 y^{-1} + V_2 y^{-2}\right), \tag{5}$$

where $\{V_0, V_1, V_2\}$ is a set of potential parameters that are related to the parameters $A_\pm$ and $\gamma$ of the differential equation (4). The solvable problems in this class include the Coulomb problem, the 3D isotropic (or 1D harmonic) oscillator, and the 1D Morse potential. The energy polynomials, $\{P_n(\varepsilon)\}$, in this class are the two-parameter Meixner-Pollaczek polynomial or the three-parameter continuous dual Hahn polynomial and their discrete versions: the Meixner, Krawtchouk or dual Hahn polynomials. However, these polynomials are well known in the appropriate physics and mathematics literature [13].

The same cannot be said about the energy polynomials in the Jacobi class to be presented next.

In the Jacobi class, the basis functions are written as follows

$$\phi_n(x)=A_n(1-y)^{\alpha}(1+y)^{\beta}P_n^{(\mu,\nu)}(y)\,, \tag{6}$$

where $y=y(x)$, $P_n^{(\mu,\nu)}(y)$ is the Jacobi polynomial of degree *n* in *y* and the normalization constant is chosen as $A_n=\sqrt{\frac{2n+\mu+\nu+1}{2^{\mu+\nu+1}}\frac{\Gamma(n+1)\Gamma(n+\mu+\nu+1)}{\Gamma(n+\mu+1)\Gamma(n+\nu+1)}}$. The dimensionless real parameters $\{\alpha,\beta,\mu,\nu\}$ are obtained by the requirements of square integrability and tridiagonalization. The dimensionless variable $y$ is a function of the physical configuration space coordinate *x* such that $-1\le y(x)\le +1$ and $\frac{dy}{dx}=\lambda(1-y)^a(1+y)^b$. The real dimensionless constants $a$ and $b$ are fixed for each choice of coordinate transformation associated with a given problem (see Appendix B in Ref. [6] for details). The basis parameters are fixed by the potential parameters and the tridiagonal requirement on the matrix representation of the wave operator. The second order linear differential wave equation associated with all solvable problems in this class has the following form in the atomic units $\hbar=M=1$

$$\left\{\left(1-y^2\right)\frac{d^2}{dy^2}-\left[a-b+y\left(a+b\right)\right]\frac{d}{dy}+\frac{A_+}{1+y}+\frac{A_-}{1-y}+A_1y\right\}\psi(y)=\gamma\,\psi(y)\,. \tag{7}$$

The corresponding potential functions in this class all have the following general form [6]

$$V(x)=(1-y)^{2a-1}(1+y)^{2b-1}\left[\frac{V_+}{1+y}+\frac{V_-}{1-y}+V_0+V_1y\right], \tag{8}$$

where $\{V_0,V_1,V_\pm\}$ is the set of potential parameters that are related to the parameters $\{A_\pm,A_1,\gamma\}$ of the differential equation (7). The solvable problems in this class include the conventional potentials (e.g., the Pöschl-Teller, Scarf, Eckart, Rosen-Morse, etc.) as well as new potentials or generalization of the conventional potentials. Table 1 is a partial list of these potentials that are exactly solvable in the TRA but not solvable using the conventional methods. The energy polynomials associated with this class of problems are the target of this study. None of them is found in the mathematics literature. We define them by their three-term recursion relations which enables us to obtain all of them analytically (albeit not in closed form) and to any desired degree starting with $P_0(\varepsilon)=1$. The three-term recursion relation is obtained from the tridiagonal matrix representation of the wave equation of the corresponding physical problem [6].

# II. The first polynomial class

The first polynomial, which we designate as $\bar{H}_n^{(\mu,\nu)}(z^{-1};\alpha,\theta)$, is a four-parameter orthogonal polynomial associated with the potential function (8). It satisfies a three-term recursion relation that could be written in the following form

$$\begin{aligned}(\cos\theta)\bar{H}_n^{(\mu,\nu)}(z^{-1};\alpha,\theta)=&\left\{z^{-1}\sin\theta\left[\left(n+\frac{\mu+\nu+1}{2}\right)^2+\alpha\right]+\frac{\nu^2-\mu^2}{(2n+\mu+\nu)(2n+\mu+\nu+2)}\right\}\bar{H}_n^{(\mu,\nu)}(z^{-1};\alpha,\theta)\\&+\frac{2(n+\mu)(n+\nu)}{(2n+\mu+\nu)(2n+\mu+\nu+1)}\bar{H}_{n-1}^{(\mu,\nu)}(z^{-1};\alpha,\theta)+\frac{2(n+1)(n+\mu+\nu+1)}{(2n+\mu+\nu+1)(2n+\mu+\nu+2)}\bar{H}_{n+1}^{(\mu,\nu)}(z^{-1};\alpha,\theta)\end{aligned} \tag{9}$$

where $n=1,2,...$, $0\le\theta\le\pi$ and $\bar{H}_0^{(\mu,\nu)}(z^{-1};\alpha,\theta)=1$. It is a polynomial of order *n* in $z^{-1}$ and in $\alpha$. In the limit $z\to\infty$, this recursion relation becomes that of the Jacobi polynomial $P_n^{(\mu,\nu)}(\cos\theta)$. The polynomial of the first kind satisfies this recursion relation with $\bar{H}_0^{(\mu,\nu)}(z^{-1};\alpha,\theta)=1$ and

$$\bar{H}_1^{(\mu,\nu)}(z^{-1};\alpha,\theta)=\frac{\mu-\nu}{2}+\frac{1}{2}\left(\mu+\nu+2\right)\left\{\cos\theta-z^{-1}\sin\theta\left[\frac{1}{4}\left(\mu+\nu+1\right)^2+\alpha\right]\right\}, \quad (10)$$

which is obtained from (9) with $n=0$ and by defining $\bar{H}_{-1}^{(\mu,\nu)}(z^{-1};\alpha,\theta)\equiv 0$. The polynomial of the second kind satisfies the same recursion relation (9) with $\bar{H}_0^{(\mu,\nu)}(z^{-1};\alpha,\theta)=1$ but $\bar{H}_1^{(\mu,\nu)}(z^{-1};\alpha,\theta)=c_0+c_1 z^{-1}$ where the linearity coefficients $c_0$ and/or $c_1$ are different from those in Eq. (10). The orthonormal version of $\bar{H}_n^{(\mu,\nu)}(z^{-1};\alpha,\theta)$ is defined as $H_n^{(\mu,\nu)}(z^{-1};\alpha,\theta)=\mathcal{A}_n\bar{H}_n^{(\mu,\nu)}(z^{-1};\alpha,\theta)$ where $\mathcal{A}_n=\sqrt{\frac{2n+\mu+\nu+1}{\mu+\nu+1}\frac{n!(\mu+\nu+1)_n}{(\mu+1)_n(\nu+1)_n}}$ and $(a)_n=a(a+1)(a+2)...(a+n-1)=\frac{\Gamma(n+a)}{\Gamma(a)}$ (see Appendix E in Ref. [6]). This polynomial has only a continuous spectrum over the whole real *z* line. This could be verified numerically by looking at the distribution of its zeros for a very large order. Figure 1 shows such distribution for a given set of values of the polynomial parameters. Whereas, Figure 2 shows that the asymptotic behavior of $H_n^{(\mu,\nu)}(z^{-1};\alpha,\theta)$ is sinusoidal and consistent with the limit given by formula (2) for $\tau=\frac{1}{2}$ and $\xi=1$. Table 2 gives the expression for *z* and the polynomial parameters in terms of the energy and potential parameters of the corresponding generalized potentials of Table 1. Consequently, if the properties of this polynomial were to be known analytically then we would have easily obtained the physical features of the corresponding quantum mechanical system. For example, if all elements in the asymptotics (2) were given explicitly, then we would have obtained the phase shift and energy spectrum in a simple and straightforward manner just by substituting the parameter map of Table 2 into Eq. (2). Additionally, the physics of the problems associated with this polynomial suggests that it should have two discrete versions. One with an infinite discrete spectrum and another that has a finite discrete spectrum [6]. These two polynomials are defined by their recursion relations, which are obtained by the replacements $\theta\to i\theta$ and $z\to iz_k$ in Eq. (9), where *k* is an integer of either finite or infinite range. Doing so, results in the following recursion relation

$$\begin{aligned}(1+\beta)\tilde{H}_n^{(\mu,\nu)}(k;\alpha,\beta)&=\left\{z_k^{-1}(1-\beta)\left[\left(n+\frac{\mu+\nu+1}{2}\right)^2+\alpha\right]+\frac{2(\nu^2-\mu^2)\sqrt{\beta}}{(2n+\mu+\nu)(2n+\mu+\nu+2)}\right\}\tilde{H}_n^{(\mu,\nu)}(k;\alpha,\beta)\\&+\frac{4(n+\mu)(n+\nu)\sqrt{\beta}}{(2n+\mu+\nu)(2n+\mu+\nu+1)}\tilde{H}_{n-1}^{(\mu,\nu)}(k;\alpha,\beta)+\frac{4(n+1)(n+\mu+\nu+1)\sqrt{\beta}}{(2n+\mu+\nu+1)(2n+\mu+\nu+2)}\tilde{H}_{n+1}^{(\mu,\nu)}(k;\alpha,\beta)\end{aligned} \quad (11)$$

where $\beta=e^{-2\theta}$ with $\theta>0$. We should note that numerical instabilities or divergences could arise in the calculation with this polynomial. The reason is seen from the recursion relation (9) where the diagonal recursion coefficient goes as $n^2$ for large *n* whereas the off diagonal coefficients go like $n^0$. For this reason (as an example), we had to choose a large value for *z* to obtain a stable and convergent evaluation of the asymptotics shown in Fig. 2.

# III. The second polynomial class

The second polynomial is a three-parameter orthogonal polynomial associated with the potential function (8) when either $V_+ = 0$ or $V_- = 0$. We designated the polynomial as $\bar{G}_n^{(\mu,\nu)}(z^2;\sigma)$ and it satisfies the following there-term recursion relation for $n = 1,2,...$

$$\begin{aligned} z^2\,\bar{G}_n^{(\mu,\nu)}(z^2;\sigma) &= \left\{\left(\sigma + B_n^2\right)\left[\frac{\mu^2-\nu^2}{(2n+\mu+\nu)(2n+\mu+\nu+2)}+1\right] - \frac{2n(n+\nu)}{2n+\mu+\nu} - \tfrac{1}{2}(\mu+1)^2\right\}\bar{G}_n^{(\mu,\nu)}(z^2;\sigma) \\ &-\left(\sigma + B_{n-1}^2\right)\frac{2(n+\mu)(n+\nu)}{(2n+\mu+\nu)(2n+\mu+\nu+1)}\bar{G}_{n-1}^{(\mu,\nu)}(z^2;\sigma) - \left(\sigma + B_n^2\right)\frac{2(n+1)(n+\mu+\nu+1)}{(2n+\mu+\nu+1)(2n+\mu+\nu+2)}\bar{G}_{n+1}^{(\mu,\nu)}(z^2;\sigma) \end{aligned} \tag{12}$$

where $B_n = \left(n + \frac{\mu+\nu}{2} + 1\right)$ and $\bar{G}_0^{(\mu,\nu)}(z^2;\sigma) = 1$. It is a polynomial of order $n$ in $z^2$. The polynomial of the first kind satisfies this recursion relation with $\bar{G}_0^{(\mu,\nu)}(z^2;\sigma) = 1$ and

$$\bar{G}_1^{(\mu,\nu)}(z^2;\sigma) = \mu + 1 - \left(\mu + \nu + 2\right)\frac{z^2 + \frac{1}{2}(\mu+1)^2}{2\left(\sigma + B_0^2\right)}. \tag{13}$$

The polynomial of the second kind satisfies the same recursion relation (12) with $\bar{G}_0^{(\mu,\nu)}(z^2;\sigma) = 1$ but $\bar{G}_1^{(\mu,\nu)}(z^2;\sigma) = c_0 + c_1 z^2$ where the linearity coefficients $\{c_0, c_1\}$ are different from those in Eq. (13). $G_n^{(\mu,\nu)}(z^2;\sigma)$ is the orthonormal version of $\bar{G}_n^{(\mu,\nu)}(z^2;\sigma)$ , which is defined by $G_n^{(\mu,\nu)}(z^2;\sigma) = \mathcal{A}_n\bar{G}_n^{(\mu,\nu)}(z^2;\sigma)$ (see Appendix E in Ref. [6]). If $\sigma$ is positive then this polynomial has only a continuous spectrum on the positive $z^2$ line. However, if $\sigma$ is negative then the spectrum is a mix of a continuous part on the positive $z^2$ line and a discrete part on the negative $z^2$ line. This could also be verified numerically by looking at the distribution of the zeros of this polynomial for a very large order. Figure 3 is a plot of such a distribution for a given set of values of polynomial parameters with negative $\sigma$ showing the continuous as well as the discrete spectrum. On the other hand, Figure 4 is a plot of the asymptotics of $G_n^{(\mu,\nu)}(z^2;\sigma)$ showing clearly the sinusoidal behavior portrayed in formula (2) where numerical analysis gives $\tau = \frac{1}{2}$ and shows that $n^{\xi} \to \ln(z)$. That is, in the asymptotic limit as $n \to \infty$

$$G_n^{(\mu,\nu)}(z^2;\sigma) \approx \frac{1}{\sqrt{n}} A(z) \times \cos\left[\ln(n)\,\theta(z) + \delta(z)\right], \tag{14}$$

where, in general, the three functions $A(z)$, $\theta(z)$ and $\delta(z)$ do also depend on the polynomial parameters $\{\sigma, \mu, \nu\}$. Table 3 gives expressions for $z^2$ and the polynomial parameters in terms of the energy and potential parameters of the corresponding physical system. Here too, if the properties of this polynomial were to be known analytically then we would have obtained the physical features of the corresponding quantum mechanical system. For example, if the asymptotics were given explicitly as in Eq. (14), we would have obtained the phase shift and energy spectrum in a simple and straightforward manner just by substituting the parameter map of Table 3 into Eq. (14). Nonetheless, the well-known energy spectrum of the potentials associated with this polynomial gives its spectrum formula as [14]

$$z_n^2 = -2\left(n + \tfrac{\nu+1}{2} - \sqrt{-\sigma}\right)^2, \tag{15}$$

where $n = 0,1,..,N$ and $N$ is the largest integer less than or equal to $\sqrt{-\sigma} - \frac{\nu+1}{2}$. Therefore, the zeros of the scattering amplitude $A(z)$ in the asymptotics (14) are at $\mathrm{i}\sqrt{-z_n^2}$ . Moreover, the known scattering states of the quantum systems associated with this polynomial (see, for example, Refs. [15-17]) give the phase shift $\delta(z)$ in the asymptotics (14) as

$$\delta(z)=\arg\Gamma\left(i\sqrt{2}z\right)-\arg\Gamma\left(\tfrac{\nu+1}{2}-\sqrt{-\sigma}+\tfrac{i}{\sqrt{2}}z\right)-\arg\Gamma\left(\tfrac{\nu+1}{2}+\sqrt{-\sigma}+\tfrac{i}{\sqrt{2}}z\right). \qquad (16)$$

Considering this result together with the spectrum formula (15) suggest that the scattering amplitude $A(z)$ in the asymptotics (14) is proportional to

$$A(z)\sim\left|\Gamma\left(i\sqrt{2}z\right)\Big/\Gamma\left(\tfrac{\nu+1}{2}-\sqrt{-\sigma}+\tfrac{i}{\sqrt{2}}z\right)\Gamma\left(\tfrac{\nu+1}{2}+\sqrt{-\sigma}+\tfrac{i}{\sqrt{2}}z\right)\right|. \qquad (17)$$

These findings also imply that the continuous weight function for this polynomial takes the following form

$$\rho(z)\propto\frac{1}{A^2(z)}\sim\left|\Gamma\left(\tfrac{\nu+1}{2}-\sqrt{-\sigma}+\tfrac{i}{\sqrt{2}}z\right)\Gamma\left(\tfrac{\nu+1}{2}+\sqrt{-\sigma}+\tfrac{i}{\sqrt{2}}z\right)\Big/\Gamma\left(i\sqrt{2}z\right)\right|^2. \qquad (18)$$

Finally, the orthogonality relation for negative $\sigma$ reads as follows

$$\int_0^\infty\rho(z)G_n^{(\mu,\nu)}(z^2;\sigma)G_m^{(\mu,\nu)}(z^2;\sigma)dz+\sum_{k=0}^{N}\omega_k G_n^{(\mu,\nu)}(z_k^2;\sigma)G_m^{(\mu,\nu)}(z_k^2;\sigma)=\delta_{n,m}, \qquad (19)$$

where $\omega_k$ is the normalized discrete weight function. For positive $\sigma$, however, only the integration part of this orthogonality survives. Moreover, the physics of the problems associated with this polynomial suggest that it has one discrete version with a finite spectrum [6]. This is defined by its recursion relations, which is obtained by the substitution $z\to iz_k$ in Eq. (12), where $k,n=0,1,..,N$ and $\sigma$ is some proper function of $N$.

# IV. Conclusion

With this short numerical expose, we hope that we did demonstrate the significance of the two polynomials $\bar{H}_n^{(\mu,\nu)}(z^{-1};\alpha,\theta)$ and $\bar{G}_n^{(\mu,\nu)}(z^2;\sigma)$ along with their discrete versions to physics. Thus, we call upon experts in the field of orthogonal polynomials to study them and derive their properties analytically.

## Table Captions:

**Table 1**: Partial list of new and/or generalized potentials that are exactly solvable using the TRA as outlined in Ref. [6].

**Table 2**: The argument and parameters of the polynomial $\bar{H}_n^{(\mu,\nu)}(z^{-1};\alpha,\theta)$ in terms of the energy and potential parameters of the corresponding generalized potential of Table 1 with $u_i = 2V_i/\lambda^2$ and $\varepsilon = 2E/\lambda^2$.

**Table 3**: The argument and parameters of the polynomial $\bar{G}_n^{(\mu,\nu)}(z^2;\sigma)$ in terms of the energy and potential parameters of the corresponding potential with $u_i = 2V_i/\lambda^2$ and $\varepsilon = 2E/\lambda^2$. The dependence of $\mu$ on the physical parameters (expected to be a simple linear relation) is yet to be derived with *N* being the number of bound states.

## Figure Captions:

**Fig. 1**: Distribution of the zeros of the polynomial $\bar{H}_N^{(\mu,\nu)}(z^{-1};\alpha,\theta)$ on the *z*-axis for a large order *N* showing that the spectrum of this polynomial is continuous over the whole real line. We took $\mu = 2$, $\nu = 3$, $\alpha = 1$, $\theta = 0.7\pi$ and $N = 500$.

**Fig. 2**: Plot of the asymptotics of $H_n^{(\mu,\nu)}(z^{-1};\alpha,\theta)$ showing clearly the sinusoidal behavior portrayed in Eq. (2) with $\tau = \frac{1}{2}$ and $\xi = 1$. We took $\mu = 2$, $\nu = 3$, $\alpha = 5$, $\theta = 0.2$ radians, $z = 10^6$ and *n* ranges from 400 to 470.

**Fig. 3**: Distribution of the zeros of the polynomial $\bar{G}_N^{(\mu,\nu)}(z^2;\sigma)$ on the $z^2$-axis for a large order *N* and for negative values of *σ*. It is evident that the spectrum in this case is a mix of a continuous part on the positive $z^2$-axis and a discrete finite part on the negative $z^2$-axis. The bottom graph is obtained by zooming in the top near the origin to show the discrete 4-point spectrum. These four points do not change if we change the polynomial order *N*. Moreover, their values and number are exactly those given by formula (15). We took $\mu = 2$, $\nu = 3$, $\sigma = -35$ and $N = 200$.

**Fig. 4**: Plot of the asymptotics of $G_n^{(\mu,\nu)}(z^2;\sigma)$ showing clearly the sinusoidal behavior depicted by Eq. (14). We took $\mu = 1$, $\nu = 2$, $\sigma = 3$, $z = 10$ and *n* ranges from 150 to 1000.

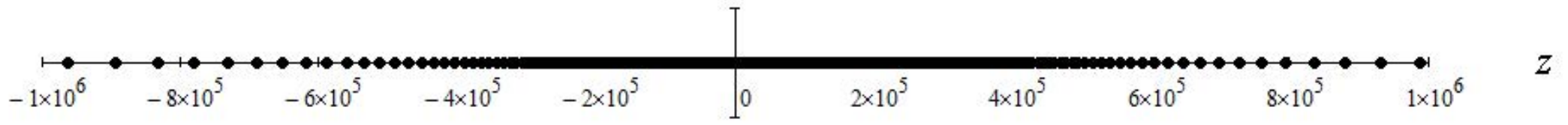

**Fig. 1**

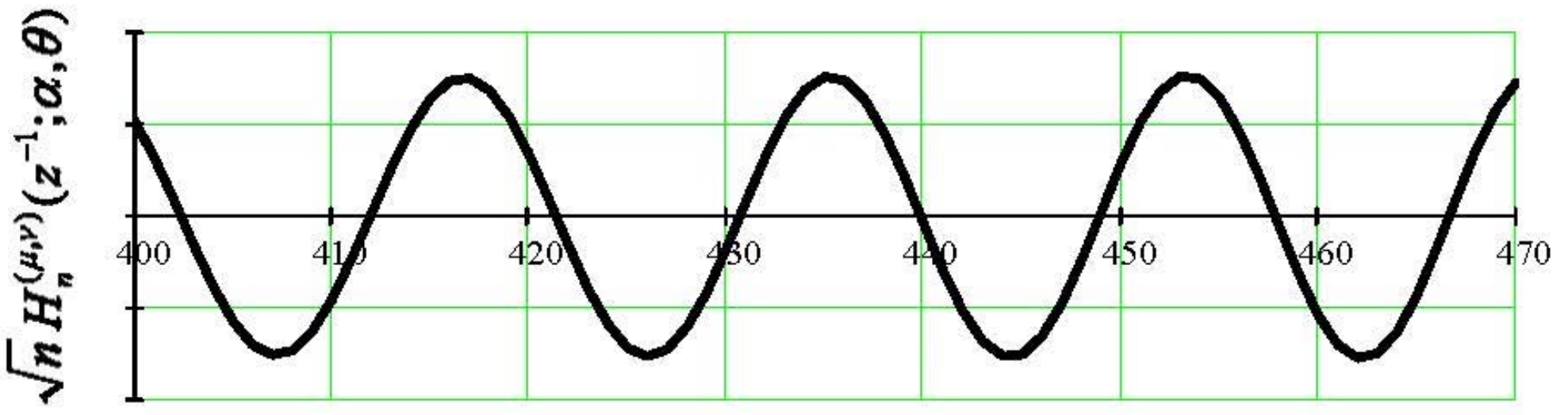

**Fig. 2**

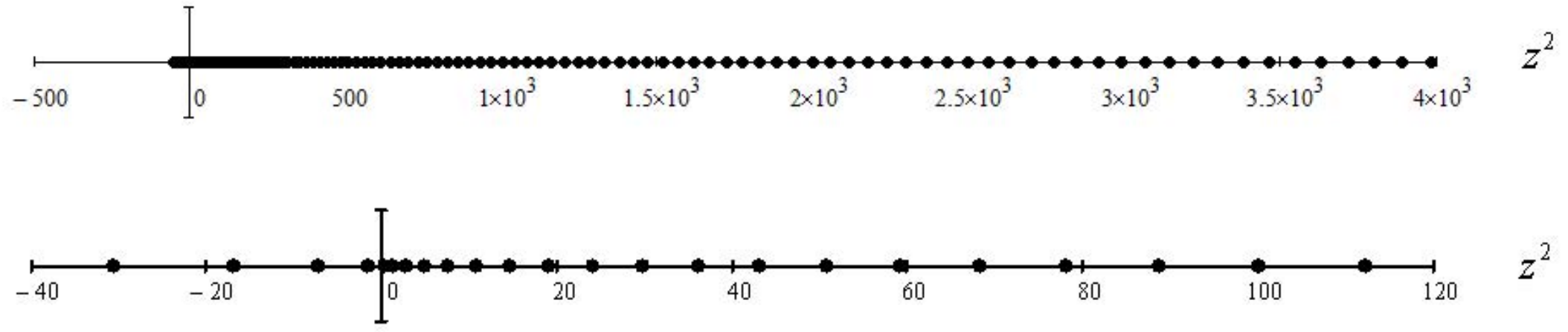

**Fig. 3**

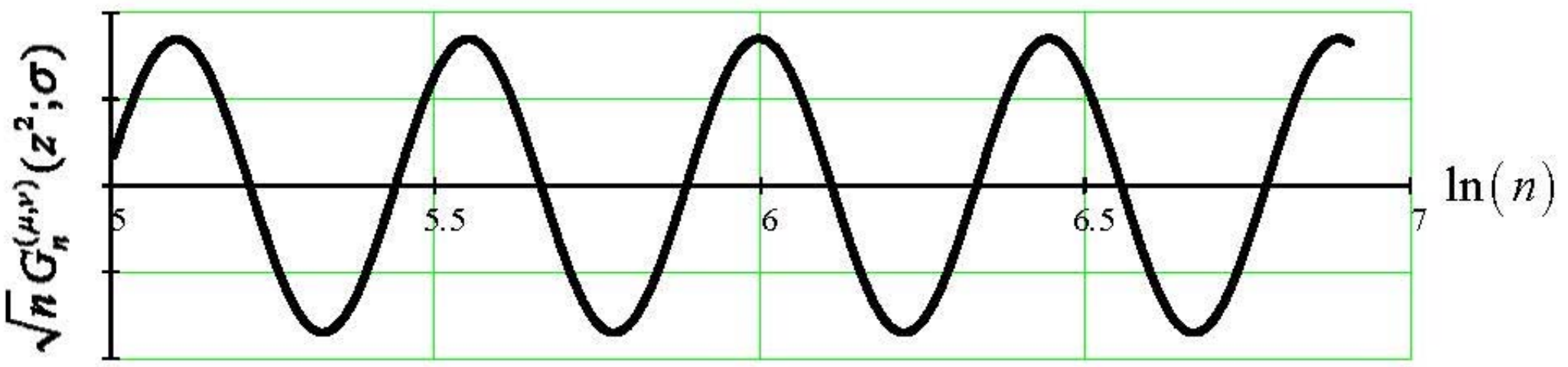

**Fig. 4**

**Table 2**

| Generalized Potential | $\cos\theta$ | $z$ | $\alpha$ | $\mu^2$ | $\nu^2$ |
|---|---|---|---|---|---|
| Trigonometric Scarf | $\frac{\varepsilon_k}{u_1}$ | $\sqrt{u_1^2-\varepsilon_k^2}$ | $u_0$ | $\frac{1}{4}+2u_-$ | $\frac{1}{4}+2u_+$ |
| New | $\frac{\varepsilon-u_1}{\varepsilon+u_1}$ | $\frac{1/2}{\sqrt{u_1\varepsilon}}$ | $u_0-u_1-\frac{1}{16}$ | $1+2u_-$ | $\frac{1}{4}+2u_+$ |
| Hyperbolic Eckart | $-\frac{u_0}{u_1}$ | $\sqrt{u_1^2-u_0^2}$ | $0$ | $-4\varepsilon$ | $1+2u_+$ |
| Hyperbolic Pöschl-Teller | $-\frac{u_0}{u_1}$ | $\frac{1}{2}\sqrt{u_1^2-u_0^2}$ | $-\frac{1}{16}$ | $-\varepsilon$ | $\frac{1}{4}+u_+$ |
| Hyperbolic Rosen-Morse | $-\frac{u_0}{u_1}$ | $\sqrt{u_1^2-u_0^2}$ | $-\frac{1}{4}$ | $-\varepsilon$ | $-\varepsilon$ |

**Table 3**

| Potential $V(x)$ | $z^2$ | $\sigma$ | $\mu^2$ | $\nu^2$ |
|---|---|---|---|---|
| $\frac{V_+}{\sinh^2(\lambda x)}+\frac{2V_0}{\cosh^2(\lambda x)}$ | $\frac{\varepsilon}{2}$ | $\frac{u_0}{2}-\frac{1}{16}$ | $f(N,u_0,u_+)$ | $\frac{1}{4}+u_+$ |
| $V_0+\frac{(V_++V_-)-(V_+-V_-)\sin(\pi x/L)}{\cos^2(\pi x/L)}$ | $-u_--\frac{1}{8}$ | $u_0-\varepsilon$ | $f(u_0,u_\pm)$ | $\frac{1}{4}+2u_+$ |
| $\frac{1}{e^{\lambda x}-1}\left[V_0+\frac{V_+/2}{1-e^{-\lambda x}}\right]$ | $2\varepsilon$ | $u_0+\varepsilon$ | $f(N,u_0,u_+)$ | $1+2u_+$ |

**Table 1**

| $y(x)$ | $V(x)$ | Generalized Potential | Bound | Scattering |
|---|---|---|---|---|
| $y(x)=\sin(\pi x/L)$<br>$-L/2 \le x \le +L/2$, $\lambda=\pi/L$ | $V_0+\frac{W_+ - W_- \sin(\pi x/L)}{\cos^2(\pi x/L)}+V_1\sin(\pi x/L)$<br>$W_\pm = V_+ \pm V_-$, $V_\pm \ge -(\pi/4L)^2$ | Trigonometric Scarf | Infinite | No |
| $y(x)=2(x/L)^2-1$<br>$0 \le x \le L$, $\lambda=2\sqrt{2}/L$ | $\frac{1/4}{1-(x/L)^2}\left[2V_0+\frac{V_+}{(x/L)^2}+\frac{V_-}{1-(x/L)^2}\right]-V_1\frac{(x/L)^2-\frac{1}{2}}{(x/L)^2-1}$<br>$V_+ \ge -1/2L^2$ $V_- \ge -2/L^2$ | New | Infinite | No |
| $y(x)=1-2e^{-\lambda x}$<br>$x \ge 0$ | $\frac{1}{e^{\lambda x}-1}\left[V_0+V_1\left(1-2e^{-\lambda x}\right)+\frac{V_+/2}{1-e^{-\lambda x}}\right]$<br>$V_+ \ge -(\lambda/2)^2$, $V_-=0$, $E \le 0$ | Hyperbolic Eckart | Finite | Yes |
| $y(x)=2\tanh^2(\lambda x)-1$<br>$x \ge 0$ | $\frac{V_+}{\sinh^2(\lambda x)}+2\frac{V_0+V_1[2\tanh^2(\lambda x)-1]}{\cosh^2(\lambda x)}$<br>$V_+ \ge -\lambda^2/8$, $V_-=0$, $E \le 0$ | Hyperbolic Pöschl-Teller | Finite | Yes |
| $y(x)=\tanh(\lambda x)$<br>$-\infty < x < +\infty$ | $\frac{V_0+V_1\tanh(\lambda x)}{\cosh^2(\lambda x)}$<br>$V_\pm = 0$, $E \le 0$ | Hyperbolic Rosen-Morse | Finite | Yes |